\documentclass[aps,prc,dlspace,twocolumn,a4paper]{revtex4}
\usepackage{amsfonts}
\usepackage{epsfig}
\usepackage{graphics}
\usepackage{color}
\newcommand{\beq}{\begin{equation}}
\newcommand{\eeq}{\end{equation}}
\newcommand{\bqa}{\begin{eqnarray}}
\newcommand{\eqa}{\end{eqnarray}}
\newcommand{\jsi}{J / \psi}

\newcommand{\sip}{\psi^\prime}
\newcommand{\eps}{\epsilon_s}

\newcommand{\jpsi}{J / \psi}

\newcommand{\pt}{p_{{}_T}}
\newcommand{\ptmin}{p_{{}_T}^{\mbox{min}}}
\newcommand{\ptmax}{p_{{}_T}^{\mbox{max}}}
\newcommand{\spt}{S(p_{{}_T})}
\newcommand{\sinpt}{ \langle S(p_{{}_T})\rangle}
\newcommand{\sincl}{\langle S^{{}^{\mbox{incl}}} \rangle}
\newcommand{\sdir}{\langle S^{{}^{\mbox{dir}}} \rangle}
\begin{document}
\title{Charmonium suppression in the presence of dissipative forces in a 
strongly coupled quark-gluon plasma}
\author{Binoy Krishna Patra$^{a}$}
\email{binoyfph@iitr.ernet.in}
\author{Vineet Agotiya$^{a}$}
\author{Vinod Chandra$^{b}$}
\affiliation{$^a$Department of Physics, Indian Institute of 
Technology Roorkee, Roorkee 247 667, India}
\affiliation{$^b$ Department of Theoretical Physics,
Tata Institute of Fundamental Research,
Homi Bhabha Road, Mumbai 400 005, India}
\date{\today}

\begin{abstract}
We study the survival of charmonium states in a strongly-coupled 
quark-gluon plasma in the presence of dissipative forces.
We consider first-order dissipative corrections to the plasma equation 
of motion in the Bjorken boost-invariant expansion with a 
strongly-coupled equation of state for QGP and
study the survival of these states with the dissociation temperatures
obtained by correcting the full Cornell potential 
not its Coulomb part alone with a dielectric function encoding
the effects of deconfined medium. We further 
explore the sensitivity of prompt and sequential suppression 
of these states to the shear viscosity-to-entropy density ratio, $\eta/s$ 
from perturbative QCD and AdS/CFT predictions.
Our results show excellent agreement with the 
recent experimental results at RHIC.
\end{abstract}

\maketitle

\section{Introduction}
One of the amazing discoveries of experimental measurements 
at RHIC is the surprising amount of collective flow exhibited
by the outgoing hadrons which is evidenced in both the single-particle
transverse momentum distribution\cite{expt} and the asymmetric azimuthal
distribution around the beam axis~\cite{adler}, quantified by the 
parameters, $v_1$ and $v_2$, respectively.
The parameter, $v_2$, in particular, was expected to be much smaller at RHIC
than it was at the lower energies of the SPS~\cite{coll}
; in fact, it is about twice as large. 
Theoretical calculations cannot predict such higher values of $v_1$ and $v_2$ 
unless the partonic cross-sections are artificially enhanced by more than an 
order of magnitude over perturbative QCD predictions \cite{molnar1}. Thus 
quark-gluon
matter created in these collisions is strongly interacting (SIQGP), unlike 
the type of
weakly interacting quark-gluon plasma (wQGP) expected to occur at very high 
temperatures on the basis of asymptotic freedom \cite{collin}. On the other 
hand, lattice
QCD calculations yield an equation of state that differs from an ideal gas
only by about 10\% once the temperature exceeds 1.5$T_c$~\cite{karsch}.
Furthermore, perfect fluid dynamics with zero shear and bulk
viscosities reproduces $v_2$ well up to transverse momenta of
order 1.5 GeV/c; however, it cannot be zero. Indeed, the
calculations within AdS/CFT suggests that $\eta \ge s/(4\pi)$. 

Charmonium ($J/\psi$) suppression has long been proposed as a signature of 
QGP formation\cite{matsui} and
has indeed been seen at SPS\cite{jpsi_sps} and RHIC 
experiments\cite{expt}. 
The heavy quark pair leading to the $J/\psi$ mesons are produced in
nucleus-nucleus collisions on a very short time-scale
$\sim 1/2m_{c}$, where $m_c$ is the mass of the charm quark. The pair
develops into the physical resonance over a formation time $\tau_\psi$
and traverses the plasma and (later) the hadronic matter
before leaving the interacting system to decay (into  a dilepton) to
be detected. This long `trek' inside the interacting system is fairly
`hazardous' for the $J/\psi$. Even before the resonance is formed
it may be absorbed by the nucleons streaming past it~\cite{GH}. By the
time the resonance is formed, the screening of the colour forces
in the plasma may be sufficient to inhibit a binding of the $c\overline{c}$
~\cite{matsui}. The resonance(s) could also be dissociated either 
by an energetic gluon~\cite{xu,gdiss} or by a comoving
hadron. A complete and deep understanding of 
the various stages of absorption is still missing and is under 
intense investigations~\cite{vogt}.
In order to disentangle these different effects, we must 
know the properties of quarkonium in 
medium and determine their dissociation temperatures.

The physics of a given quarkonium state is encoded in its spectral 
function\cite{Iid06,Jak07}.
So following how the spectral function changes with temperature can give us
a theoretical insight to the temperature-dependence of
quarkonium properties. There are mainly 
two lines of theoretical studies to determine quarkonium spectral functions:
the first one is the potential 
models~\cite{potential2,Shu04,Won05,Cab06} which
have been widely used to study quarkonium, but their applicability at 
finite temperature is still under scrutiny and the second one is the 
lattice QCD\cite{lat_disso} 
which provides the most straightforward way to determine
spectral functions, but the results suffer from discretization effects
and statistical errors, and thus are still inconclusive.
These two approaches show poor matching between their predictions
because of the uncertainties
coming from a variety of sources.
None of the approaches give a complete framework to study the
properties of quarkonia states at finite temperature.
However, some degree of qualitative agreement had been achieved
for the S-wave correlators. The finding was somehow ambiguous for the
P-wave correlators and the temperature dependence of the potential model
was even qualitatively different from the lattice one.
Refinement in the computations of the spectral functions have recently
been done by incorporating the zero modes both in the S- and P-channels
\cite{Ume07,Alb08}.
It was shown that, these contributions cure most of the previously observed
discrepancies with lattice calculations.
This supports the use of potential models at finite
temperature as an important tool to complement lattice studies.

It is now pertinent to ask how a quarkonium behaves
in strongly interacting QGP unlike in wQGP. The propagation of 
charm quark in SIQGP is 
different from wQGP because in SIQGP there 
will be a rapid equilibration due to the multiple momentum exchange
in the momentum space but in the coordinate space, the motion is 
diffusive in nature and slower\cite{shur}. 
Therefore, the usual picture of charmoniun suppression\cite{matsui} 
may not be true in SIQGP.
Moreover, in the RHIC era (small $\mu_{B}$), it has been 
confirmed from the recent studies that the transition from 
nuclear matter to QGP is not a phase transition, rather a 
crossover\cite{phaseT}. 
It is then reasonable to assume that the string-tension
does not vanish abruptly at the deconfinement point, so 
we may expect presence of non-perturbative effects 
such as non-zero string tension in the deconfined phase and one
should study its effects on heavy quark potential even above $T_c$.
This is indeed compatible with the strongly interacting nature 
of QGP where large number of non-perturbative effects are expected. 
This issue, usually overlooked in the
literature where only a screened Coulomb potential was 
assumed above $T_c$ and the linear/string term was assumed zero,
was certainly worth of investigation.
This was exactly we had done recently~\cite{prc-vinet} where
a medium-modified heavy quark potential was derived
by correcting the full Cornell potential
not its Coulomb part alone with
a dielectric function encoding the effects of the deconfined medium.
We found that this approach led to a long-range Coulomb potential
with an (reduced) effective charge in addition to
the usual Debye-screened form. With such an effective potential, 
we investigated the
effects of perturbative and non-perturbative contributions
to the Debye masses on the dissociation of quarkonium states.
Finally we determined the binding energies and dissociation temperatures 
of the ground and the first excited states of charmonium and bottomonium 
spectra~\cite{prc-vinet,plb-vinet}.

   Now let us consider a central collision in a nucleus–
nucleus collision, which results in the formation of quark–
gluon plasma at initial time $\tau_i$. 
Let us concentrate on the region of energy density greater than 
a characteristic energy density, say, screening energy density 
$\epsilon_s$ which
encloses the plasma which is dense enough to cause the
melting of a particular state of quarkonium. 
We assume the plasma to cool, according to Bjorken's boost 
invariant (longitudinal) hydrodynamics. 
If the plasma has cooled to a energy density 
less than $\epsilon_s$, the $c \bar c$ pair would escape
and quarkonium would be formed. If however, the energy
density is still larger than $\epsilon_s$, the resonance 
will not form and we shall have a quarkonium suppression.
It is easy to see that the $\pt$ dependence of the survival
probability will depend on how rapidly the plasma cools.
If the initial energy density is sufficiently high, the plasma
will take longer to cool and only the pairs with very high
$\pt$ will escape. If however the plasma cools rapidly, then
even pairs with moderate $\pt$ will escape. 
There are many attempts~\cite{chu,dpal,compare} where 
the above picture of plasma expansion dynamics had been
employed to study charmonium suppression, but in their works following
points are missing: a) the effects of dissipative forces was not 
included in hydrodynamic expansion. 
The presence of dissipative term in the energy-momentum tensor~\cite{rud}
causes the plasma to expand slowly and the system
will take longer to reach $\epsilon_s$ 
(screening time $\tau_s$) and only the pairs with very high
$\pt$ will escape resulting overall more suppression. 
b) The EoS employed in their works was either 
ideal or bag model EoS which was then directly used to calculate 
two crucial factors in quantifying the suppression in 
relativistic heavy ion collisions: the energy density $\epsilon_s$ 
corresponding to the dissociation temperature ($T_D$) and 
the time elapsed by the system ($\tau_s$) to reach
$\epsilon_s$ by the expansion (cooling) of the system. As we now know the 
matter formed at RHIC is far from its
ideal limit even at $T \geq T_c$ so the EoS employed in their works was not 
reliable to capture the properties of strongly-interacting matter.
This is evident from the substantial difference in the numerical values
of the crucial quantities calculated in the ideal
and strongly-interacting equation of state (in the Tables II-V).

The main motivation of this article is to remedy
the above shortcomings in multifold respects:
i) First we use an appropriate equation of state (EoS) which should
reproduce the 
lattice results verifying the strongly-interacting nature of QGP.
This also explains the non-ideal
behaviour of QGP through the speed of sound which will play an important
role in the expansion dynamics.
ii) Then we study hydrodynamic boost-invariant Bjorken expansion 
in $(1+1)$ dimension with the EoS discussed in i) as an input.
In addition we explore the effects of dissipative terms on the
hydrodynamic expansion by considering the shear viscosity $\eta$ 
up to first-order in the stress-tensor. Basically we consider 
the ratio of the shear viscosity-to-entropy density ($\eta/s$)
as $1/{4\pi}$ and 0.3 which was predicted from the 
AdS/CFT correspondence~\cite{ads_dtson} and perturbative QCD~\cite{pqcd}
, respectively.
We also consider ideal hydrodynamics with $\eta/s=0$ for 
the sake of comparison. 
iii) The most important point is to know the properties of quarkonium in 
the medium. For that we need to know the correct form of medium-modified 
potential to determine the dissociation temperatures of charmonium states
in a static, thermal medium. For this purpose we follow our recent work
where medium modified potential~\cite{prc-vinet,plb-vinet}
was obtained by correcting the full form of the Cornell potential through a
dielectric function embodying the medium effects. iv) Finally we study 
the survival of charmonium states
with all the refinements discussed above.
The extent of suppression will be decided by a competition
among the dissociation scale, the transverse momentum of $J /\psi$'s and 
the rate of expansion, making it
sensitive to such details as the speed of sound~\cite{chu,dpal}
through the EoS. Thus a study of the survival of $J/\psi$ is poised to provide
a wealth of information about the dissociation mechanism of $c \bar c$
states in a hot QCD medium, the evolution of the plasma and 
its properties.

The paper is organized as follows. In Sec.II.A, we briefly describe
the strongly-interacting equation of state developed by 
Bannur~\cite{bannur1,ba1} and determine the pressure,
energy density and the speed of sound etc.
In Sec.II.B, we then use the aforesaid equation of state
to study boost-invariant (1+1) dimensional longitudinal expansion
in the presence of viscous forces. In Sec.III.A, we study the dissociation
phenomenon of $J/\psi$ by studying the in-medium modifications
to the heavy quark potential and its suppression in a
longitudinally  expanding QGP in Sec.III.B. 
Results are presented in Sec.IV with a 
discussion on important aspects of our approach. Finally in Sec.V, we 
present the conclusions and future prospects of the present work.

\section{Strongly-interacting QGP and Longitudinal expansion of QGP}
In this section, we first discuss the equation of state for 
SIQGP. We shall then discuss 
hydrodynamical expansion in the presence of viscous forces
with the SIQGP EoS as an input.

\subsection{Equation of state}
The equation of state for the quark matter produced in RHIC 
is a very important observable and
the properties of the matter are sensitive to it. 
The expansion of QGP is quite sensitive to the 
equation of state(EoS) through the speed of sound.
Both experimental~\cite{gu1} and theoretical (lattice) results~\cite{ka1}
show that matter formed near or above $T_c$ is
non-ideal. There have been many attempts to explain
such a matter using various models such as bag models, other
confinement models~\cite{ba1}, quasi particle models~\cite{pe1}, strongly 
interacting quark gluon plasma (sQGP)~\cite{sh1} etc.

Here we propose that the QGP near $T_c$ is in fact 
what is called
strongly coupled plasma~\cite{bannur1}, widely studied in QED plasma
where the plasma parameter, $\Gamma$, defined as the ratio of the average
potential energy to the average kinetic energy of the particles, is of the
order of 1 or larger. An extensive study  of strong-coupled plasma in QED
with proper modifications to include colour degrees of
freedom and the strong running coupling constant 
gives an expression for the energy density
as a function of the plasma parameter, $\Gamma$
\begin{equation}
\label{relativistic}
\epsilon=(2.7+u_{ex}(\Gamma))nT \quad ,
\end{equation}
where the first term corresponds to the ideal EoS and the second 
term, $u_{\rm{ex}}(\Gamma)$, 
gives the non-ideal (or excess) contribution to EoS as
\begin{equation}
\label{abe}
u_{ex}(\Gamma)=-\frac{\sqrt{3}}{2}\Gamma^{3/2}
\end{equation}
So, the scaled energy density (in terms of Stefan-Boltzmann limit)
is given by
\begin{equation}
\label{egamma1}
e(\Gamma)\equiv\frac{\epsilon}{\epsilon_{SB}}
=1-\frac{1}{2.7}\frac{\sqrt{3}}{2}\Gamma^{3/2} \quad ,
\end{equation}
where
${\epsilon_{SB}}\equiv 3{a_f}T^4$
with degrees of freedom ${a_f}\equiv(16+21{n_f/2}){{\pi^2}/90}$
and using the
relation, $\epsilon= T \frac{\partial p}{\partial T} -p$,
we get the pressure
\begin{equation}
\label{pressure}
\frac{p}{T^4}= \left(\frac{p_0}{T_0}+3{a_f}T^4 \int^{T_0}_{T}
d\tau{\tau^2}e(\Gamma(\tau)) \right) /T^3 \quad, 
\end{equation}
where $p_0$ is the pressure at some reference temperature $T_0$.
If we calculate $p(T)/T^4$ versus $T$ for pure gauge, 2-flavor 
and 3-flavor QGP, a surprisingly good fit with the lattice data was 
found~\cite{bannur1,ba1}.

Another important observable, the speed of sound which appears in 
the equation of motion for the expansion through the relation,
\begin{equation}
\label{cs2}
c_s^2=\frac{\partial p}{\partial \epsilon}.
\end{equation}
It is found that strongly-interacting matter created 
at RHIC has a signature of  non-zero interaction measure, 
$\Delta=\epsilon-3p$ which measures the deviation of the EoS from its 
ideal counterpart. This quantity has been extensively
studied in lattice QCD~\cite{leos} which shows that 
$\Delta\ge 0$ and asymptotically approaches to the ideal value (zero). 
This implies that $c_s^2 \le 1/3$ at $T=T_c$ or even much above $T_c$.
This is exactly found in Bannur's model~\cite {bannur1} 
where $c_s^2$ is very much less than its ideal limit around 
$T_c$ and increases towards its ideal value.
In view of the excellent agreement with lattice simulations
the above phenomenological EoS is a right choice for the 
strongly-interacting matter possibly formed at RHIC to calculate the
thermodynamical quantities {\em viz.} screening energy density ($\epsilon_s$),
the speed of sound etc. and also to study 
the hydrodynamical expansion of plasma.

\subsection{Longitudinal expansion in the presence of dissipative forces}
The energy momentum tensor of the plasma in the absence of dissipative 
forces is written as:
\begin{equation}
T^{\mu\nu}= (\epsilon+p)u^\mu u^\nu + g^{\mu \nu} p~.
\label{tmu}
\end{equation}
If the effects of viscosity are neglected, conservation 
of energy-momentum tensor leads to
\begin{equation}
\label{eom}
\partial_\mu T^{\mu\nu}= 0~.
\end{equation}
We now consider the Bjorken boost invariant longitudinal flow in (1+1) 
dimension where the equation (\ref{eom}) simplifies to 
\begin{equation}
\partial_\tau \epsilon=-\frac{\epsilon+p}{\tau}.
\label{bj}
\end{equation}
The effect of the speed of sound $c_s^2$ is seen immediately through the
EoS: $p=c_s^2 \epsilon$ as
\begin{eqnarray}
\label{eq0}
\epsilon(\tau) \tau^{1+c_s^2}&=& \epsilon(\tau_i)\tau_i^{1+c_s^2}
=\mbox{const.}
\end{eqnarray}
With the condition $\epsilon\sim T^4$ and taking the speed of sound
$c_s^2=1/3$, the above equation translates 
to the cooling law:
\begin{eqnarray}
\label{cool_id}
T^3\tau=T_i^3 \tau_i ~,
\end{eqnarray}
where $T_i$ is the initial temperature and 
$\epsilon(\tau_i)$ is the initial energy density.

Now, we study the correction to the equation of motion (\ref{eom})
in the presence of dissipative forces in the form
of shear viscosity, $\eta$ which is itself a function of temperature.
The hydrodynamic 
evolution preserves the boost-invariance, 
which in ultrarelativistic heavy ion collisions is expected to be 
realized near mid-rapidity~\cite{Bjorken}. 
In the presence of viscous forces the energy-momentum tensor is written as,
\begin{equation}
T^{\mu\nu}= (\epsilon+p)u^\mu u^\nu + g^{\mu \nu} p +\pi^{\mu\nu},
\label{tmun}
\end{equation}
where $\pi^{\mu\nu}$ is the stress-energy tensor. Dissipative corrections
to the first-order in the gradient expansion are recovered by setting the
relaxation time to zero. This leads to:
\beq
\pi^{\mu\nu}=\eta \langle \nabla^\mu u^\nu \rangle, 
\eeq
where $\eta$ is the co-efficient of the shear viscosity and 
$ \langle \nabla^\mu u^\nu \rangle$ is the symmetrized velocity gradient. 

In (1+1) dimensional Bjorken expansion in the first-order
dissipative hydrodynamics, only one component $\pi^{\eta \eta}$ of the 
viscous stress tensor is non-zero.
In this case the equation of motion reads,
\beq
\label{eqm}
\partial_\tau \epsilon =-\frac{\epsilon+p}{\tau}+\frac{4 \eta}
{3 \tau^2}\, .
\label{3.15}
\eeq
The solution of the above equation in the case of constant value of 
$\eta/s$ is known analytically ~\cite{Kouno,Muronga}
and is given by,
\beq
T(\tau)=T_i \left(\frac{\tau_i}{\tau}\right)^{1/3}\left[1+\frac{2\eta}{3 s
 \tau_i T_i}\left(1-\left(\frac{\tau_i}{\tau}\right)^{2/3}\right)\right].
 \label{3.16}
\eeq
The first term in the RHS is the same as in the case of 
zeroth-order (non-viscous) hydrodynamics and the second term is the 
correction arising from constant $\eta/s$ which causes the system
to expand slowly compared to the perfect fluid $\eta=0$ (\ref{cool_id}).
It is important to note 
that temperature will be maximum after a span of time
\begin{equation}
\tau_{\rm max} = \tau_i \left( \frac{1}{3} + \frac{s}{\eta} 
\frac{\tau_i\, T_i}{2} \right)^{-3/2}\, .
\label{3.17}
\end{equation}

For times $\tau > \tau_{\rm max}$, the temperature decreases with time, 
as expected for matter undergoing expansion. For early times 
$\tau < \tau_{\rm max}$, however,
the solution shows an unphysical reheating~\cite{wd} which will
not be discussed here.

We shall now study the solution of Eq.(\ref{3.15}) for the
strongly-interacting equation of state discussed in Sec.II.A
where $c^2_s$ is always less than $1/3$ and is a function 
of temperature. The solution of Eq.(\ref{eqm}) is obtained as,
\begin{eqnarray}
\label{eqs1}
\epsilon(\tau) \tau^{(1+c_s^2)}+ \frac{4a}{3{\tilde{\tau}}^2}
\tau^{(1+c_s^2)}
&=&\epsilon(\tau_i)\tau_i^{(1+c_s^2)}
+\frac{4a}{3 {\tilde{\tau_i}}^2} \nonumber\\
&=&{{\mbox{const}}}~,
\end{eqnarray}
where $a =\left(\frac{\eta}{s}\right)T^3_i \tau_i$ and 
${\tilde{\tau}}^2$ and ${\tilde{\tau}}_i^2$ are given by
$(1-c_s^2)\tau^2$ and $(1-c_s^2)\tau_i^2$, respectively.
The first term 
in both LHS and RHS accounts for the contributions coming from the 
zeroth-order expansion and the second term is the first-order 
viscous corrections (In other words,
Eq.(\ref{eqs1}) reduces  to Eq.(\ref{eq0})) for $\eta=0$.)

In our work we consider three values of the shear viscosity-to-entropy
density ratio to see the effects of nonzero 
values of the shear viscosity on the expansion. The first one is 
from perturbative QCD~\cite{pqcd} calculations where $\eta/s$ is 
=0.3 near $T\sim T_c-2T_c$. The second one is from AdS/CFT 
studies~\cite{ads_dtson} where $\eta/s=1/{4\pi}$ ($\sim 0.08$).
Finally we consider $\eta/s$=0 (for the ideal fluid) for the sake of 
comparison. We shall employ Eq.(\ref{eqs1}) to study the 
charmonium suppression in a strongly interacting QCD medium formed
in a ultra-relativistic heavy-ion collisions in the next section.

\section{Suppression of $J/\psi$ in a longitudinally expanding plasma}
Let us now turn our attention to the most important quantity, 
{\it viz.}, the $J/\psi$ survival probability.
Recall that, to study the fate of charmonium in SIQGP, we need to consider  
(i) the equation of state for SIQGP, (ii) the effects of dissipative forces 
(shear viscosity) on the expansion, and (iii) an appropriate 
criterion for the dissociation of charmoniun in hot QCD medium. 
The first two issues have already been 
discussed in the previous section.
Here, we first discuss the dissociation of charmonium 
states by giving a brief derivation of medium-modified
potential in Sec.III.A. We also discuss the nature of dissociation
by examining the effects of perturbative
and non-perturbative terms in the Debye masses. Finally we proceed to 
derive the survival probability of $J/\psi$ in an expanding SIQGP in
Sec.III.B.

\subsection{Dissociation of charmonium in SIQGP}
The large distance property of the heavy quark
interaction is important for our understanding of the bulk properties of
the QCD plasma phase, {\em e.g.} the screening property of the quark gluon
plasma \cite{Kac1,Kac2}, the equation of state
\cite{Bein,kars} and the order parameter (Polyakov loop)
\cite{Kac3,Kac4,Dum1,Dum2}. In all
of these studies deviations from perturbative calculations and the ideal gas
behavior are expected and were indeed found at temperatures which are 
moderately larger than the deconfinement temperature. This calls for
quantitative non-perturbative calculations.
Recent findings~\cite{phaseT}indicate the phase transition in full QCD 
appears as an crossover
rather than a `true' phase transition with the related singularities in
thermodynamic observables.
In light of this finding, one can not simply 
ignore the effects of string tension between the quark-antiquark pairs
beyond $T_c$. This is indeed a very important effect which needs
to be incorporated while setting up the criterion for the 
dissociation of quarkonia in QGP.

In our recent works\cite{prc-vinet,plb-vinet}, this issue
has successfully been addressed 
for the dissociation of quarkonium in 
QGP. In these works, an appropriate form of the medium modified  
inter-quark potential have been obtained. 
The authors have considered the Cornell form of the 
potential: $V(r)=-\alpha/r +\sigma r$ and correct its
Fourier transform (FT) $\tilde{V} (k)$ as 
\begin{equation}
\label{eq3}
\tilde{V}(k)=\frac{V(k)}{\epsilon(k)} \quad ,
\end{equation}
where $\epsilon(k)$ is the dielectric permittivity given in terms
of the static limit of the longitudinal part of the gluon
self-energy\cite{schneider}:
\begin{eqnarray}
\label{eqn4}
\epsilon(k)=\left(1+\frac{ \Pi_L (0,k,T)}{k^2}\right)\equiv
\left( 1+ \frac{m_D^2}{k^2} \right)~,
\end{eqnarray}
and $V(k)$ is the Fourier transform (FT) of
the Cornell potential:
\begin{equation}
\label{eqn5}
{\bf V}(k)=-\sqrt{(2/\pi)} \frac{\alpha}{k^2}-\frac{4\sigma}{\sqrt{2}\pi k^4}.
\end{equation}
Substituting Eqs.(\ref{eqn4}) and (\ref{eqn5}) into (\ref{eq3})
and then evaluating its inverse FT
one obtains the r-dependence of the medium modified
potential~\cite{prc-vinet,plb-vinet},
\begin{eqnarray}
\label{full}
{\bf V}(r)&=&(\frac{2\sigma}{m^2_D}-\alpha)\frac{\exp{(-m_Dr)}}{r}\nonumber\\
&-&\frac{2\sigma}{m^2_Dr}+\frac{2\sigma}{m_D}-\alpha m_D~,
\end{eqnarray}
where the constant terms
are introduced  to yield the correct limit of $V(r,T)$ as $T\rightarrow 0$.
Such terms could arise
naturally from the basic computations of real time static potential
in hot QCD\cite{const1} and from the real and imaginary time
correlators in a thermal QCD medium\cite{const2}.
The medium modified form of the potential thus obtained has an additional 
long range Coulomb term with an (reduced) effective charge in addition
to the conventional 
Yukawa term. A new picture of quarkonium suppression has emerged out 
of these studies where both the range and the charge of the potential
get screened due to in-medium modifications. 
The binding energies and dissociation temperatures for various 
quarkonium states have been determined by solving the Schr\"odinger
equation numerically with the potential (\ref{full}).
We considered three possible forms of the
Debye masses ($m_D$) in the potential:
the leading-order term in QCD coupling
($m_D^{\mbox{LO}}$) which is a perturbative result~\cite{shur}, the 
non-perturbative 
corrections to the leading-order ($m_D^{\rm{NP}}$)~\cite{kajantie} and the 
lattice parametrized form ($m_D^L=1.4 m_D^{\mbox{LO}}$)~\cite{mocsy_prl} to examine the effects of
perturbative and nonperturbative contributions on the dissociation
mechanism. The upper limit of the dissociation temperature of a particular
state is obtained when its binding energy is of the order of
temperature of the system: $E_{\rm{bin}} \approx T$ 
so that the state becomes feebly bound
and thermal fluctuations can excite it to the continuum.
The perturbative result for the Debye mass in the potential 
gives rise higher values (in Table II) of the dissociation
temperatures compared to the spectral function 
studies~\cite{mocsy_prl, mocsy_prd} calculated in a potential model
but smaller than the results from the analysis of lattice
(temporal) correlators~\cite{Dat04} (in Table IV).
On the other hand, when we use the lattice parametrized form 
(by fitting the lattice free-energy for color-singlet states)
for the Debye mass, the dissociation temperatures 
(given in Table III) exactly matches (discussed in
in details in Ref.~\cite{prc-vinet}) 
with the results of the potential model calculations~\cite{mocsy_prl, mocsy_prd}
but much smaller than the lattice studies~\cite{Dat04} (Table IV).
Finally the non-perturbative corrections of ${\cal O}(g^2T)$ and 
${\cal O}(g^3T)$ to the perturbative result~\cite{kajantie} 
in Debye mass give unrealistically smaller values which will
not be considered in the present work.

In addition, we also take the advantage
to demonstrate the flavor dependence of the dissociation
process. Hence we determined the dissociation 
temperatures for the 2-flavor QGP (Table V) with
the Debye mass only in the leading-order term in strong coupling.
The dissociation temperatures are found to be higher than 
the 3-flavor case (Table II). However the detailed study of the
flavor dependence was given in our earlier works~\cite{prc-vinet,plb-vinet}. 
In the present analysis, we have used the four sets
of the dissociation temperatures (Tables II-V) to study the centrality 
dependence of survival probability of charmonium as a function of 
number of participants $N_{\mbox{Part}}$ 
in an expanding (strongly interacting) QCD medium.

\subsection{Survival probability}
We now have all the ingredients to write down the survival probability and 
we closely follow Chu and Matsui for this.
In the work of Chu and Matsui~\cite{chu},
the $p_T$ dependence of the survival probability of
charmonium was studied by choosing the speed of sound $c_s^2=1/3$ (ideal EoS) 
and the extreme value $c_s^2=0$.
This work was further generalized by invoking the 
various parameters for Au-Au collisions at RHIC in \cite{dpal}
to include the effects of realistic EoS in 
an adhoc manner by simply choosing a lower value of $c_s^2=1/5$.
Instead of taking arbitrary values of $c_s^2$ we tabulated 
the values of $c_s^2$ in Tables II-V corresponding to the 
dissociation temperatures 
calculated from Bannur model~\cite{bannur1} where the values of
$c_s^2$ match perfectly with the lattice results.  
Moreover, in the light of recent experimental finding from RHIC,
one can not ignore the viscous effects while studying 
charmoniun suppression. Here, we address these issues.
Note that in our analysis, we follow the idea of Chu and 
Matsui\cite{chu} but with the more appropriate criteria for 
the charmonium dissociation and the dissipative hydrodynamic 
expansion of the plasma.

Let us take a simple parametrization for the initial energy-density profile on
any transverse plane :
\begin{equation}
\epsilon(\tau_i,r)=\epsilon_i A_{{}_T}(r); A_{{}_T}(r) =\left(1-
\frac{r^2}{R_T^2}\right)^{\beta} \theta(R_T-r)
\end{equation}
where $r$ is the transverse co-ordinate and $R_T$ is the transverse 
radius of the nucleus.
One can define an average energy density $<\epsilon_i>$ as
\begin{equation}
\pi R_T^2 <\epsilon_i>=\int 2\pi\, r \, dr \epsilon(r)
\end{equation}
so that 
\begin{equation}
\epsilon_i=(1+\beta)  \langle \epsilon_i \rangle.
\end{equation} 
We have taken $\beta=1$,  which will reflect the proportionality of 
the deposited energy  to the nuclear thickness. The average energy-density 
is obtained from the Bjorken formula:
\begin{equation}
\langle \epsilon_i \rangle =\frac{1}{\pi R_T^2 \tau_i}\,\frac{dE_T}{dy}
\end{equation}
where $E_T$ is the transverse energy deposited in the collision. 
The appropriate characterization of kinematic
quantities in Au+Au collisions is presented in Table I.
\begin{table}[tbp]
\caption{Kinematic characterization of Au$+$Au collisions at 
RHIC~\cite{expt}}
\begin{center}
\begin{tabular}{lll|lll}
\hline Nuclei & $\sqrt{s_{NN}}$ & $N_{\mbox{part}}$~ &~ $<\epsilon>_i$~
~ &~ $R_T$ \\ 
& (GeV) & ~ &~ (GeV/fm$^3$)~ &~ (fm)\\ \hline
         &          & 22.0 & 5.86 & 3.45 \\
         &          & 30.2 & 7.92 & 3.61 \\
         &          & 40.2 & 10.14& 3.79 \\
         &          & 52.5 & 12.76& 3.96 \\
         &          & 66.7 & 15.69& 4.16 \\
         &          & 83.3 & 18.58& 4.37 \\
Au$+$Au  & 200  & 103.0& 21.36& 4.61 \\
         &          & 125.0& 24.38& 4.85 \\
         &          & 151.0& 27.37& 5.12 \\
         &          & 181.0& 30.52& 5.38 \\
         &          & 215.0& 34.17& 5.64 \\
         &          & 254.0& 37.39& 5.97 \\
         &          & 300.0& 41.08& 6.31 \\
         &          & 353.0& 45.09& 6.68 \\\hline
\end{tabular}
\end{center}
\end{table}

The time $\tau_s$ when the energy density drops to the screening energy
density $\epsilon_s$ is estimated from Eq.({\ref{eqs1}}) as
\begin{eqnarray}
\label{taus}
\tau_s(r)=\tau_i {\bigg[ \frac{\epsilon_i(r)-
\frac{4a}{3{\tilde{\tau}}_i^2}}{\epsilon_s-\frac{4a}{3{\tilde{\tau}}_s^2}}
\bigg]}^{1/1+c_s^2}
\end{eqnarray}
where $\epsilon_i(r)=\epsilon(\tau_i,r)$ and 
${\tilde{\tau}}_s^2$ is $(1-c_s^2)\tau_s^2$.
The critical radius $r_s$, is seen to mark the boundary of the region 
where the quarkonium formation is suppressed, can be obtained by
equating the duration of screening
$\tau_s(r)$ to the formation time $t_F=\gamma \tau_F$ for the quarkonium
in the plasma frame and is given by:
\begin{eqnarray}
\label{rs}
r_s= R_T { \left( 1- A \right)}^{\frac{1}{2}} \theta \left( 1-A \right)~,
\end{eqnarray}
where $A$ is given by
\begin{eqnarray}
A &=& \bigg[  \bigg( \frac{\epsilon_s}{\epsilon_i} \bigg)
\bigg(\frac{t_F}{\tau_i} \bigg)^{1+c_s^2} \bigg. \nonumber\\
&+& \bigg. \frac{1}{\epsilon_i} {\bigg( \frac{t_F}{\tau_i} \bigg)}^{(1+c_s^2)}
\frac{4a}{3{\tilde{t}}_F^2} 
+\frac{1}{\epsilon_i} 
\frac{4a}{3{\tilde{\tau}}_i^2} \bigg]^{1/\beta}
\end{eqnarray}
with ${\tilde{t}}_F^2= (1-c_s^2)t_F^2$.
The quark-pair will escape the screening region (and form quarkonium) 
if its position $\mathbf{r}$ and 
transverse momentum $\mathbf{p}_T$ are such that
\begin{equation}
\left| \mathbf{r}+\tau_F \mathbf{p}_T/M\right| \geq r_s.
\end{equation}
Thus, if $\phi$ is the angle between the vectors $\mathbf{r}$ and 
$\mathbf{p}_T$,
 then 
\begin{equation}
\cos \phi\,\geq\,\left[(r_s^2-r^2)\,M-\tau_F^2\,p_T^2/M\right]/
\left[2\,r\,\tau_F\,p_T\right],
\label{phi}
\end{equation}
which leads to a range of values of $\phi$ when the quarkonium would
escape. Now we can write for the survival probability of the quarkonium:
\begin{eqnarray}
S(p_T)&=&\left[\int_0^{R_T} \, r \, dr \int_{-\phi_{\mbox{max}}}
^{+\phi_{\mbox{max}}}\,
d\phi\, P(\mathbf{r},\mathbf{p}_T)\right]/\nonumber\\&&
\left[2\pi \int_0^{R_T} \, r\, dr\, P(\mathbf{r},\mathbf{p}_T)\right],
\label{spt}
\end{eqnarray}
where $\phi_{\mbox{max}}$ is the maximum positive angle 
($0\leq \phi \leq \pi$)
allowed by Eq.(\ref{phi}):
\begin{equation}
\phi_{\mbox{max}}=\left\{ \begin{array}{ll}
\pi & \mbox{if $y\leq -1$}\\
\cos^{-1} |y| & \mbox{if $-1 < y < 1$}\\
 0          & \mbox{if $y \geq 1$}
 \end{array}
 \right .,
\end{equation}
where
\begin{equation}
y= \left[(r_s^2-r^2)\,M-\tau_F^2\,p_T^2/M\right]/
\left[2\,r\,\tau_F\,p_T\right],
\end{equation}
and $P$ is the probability for the quark-pair production at
($\mathbf{r}$, $\mathbf{p}_T$), in a hard collision which
may be factored out as 
\begin{equation}
P(\mathbf{r},\mathbf{p}_T)=f(r)g(p_T),
\end{equation}
where we take the profile function f(r) as
\begin{equation}
f(r)\propto \left[ 1-\frac{r^2}{R_T^2}\right]^\alpha \theta(R_T-r)
\end{equation}
with $\alpha=1/2$. 

Often experimental measurement of survival probability at a given
number of participants ($N_{{}_{\rm part}}$) or rapidity ($y$)
is reported in terms of the $\pt$-integrated yield ratio 
(nuclear modification factor) over the range $\ptmin \le \pt \le \ptmax$
whose theoretical expression would be
\begin{equation}
\sinpt  = 
\frac{\int_{\ptmin}^{\ptmax}  d \pt S(\pt)}
{\int_{\ptmin}^{\ptmax} d \pt}
\end{equation}
The production of  $J/\psi$ mesons
in hadronic reactions occurs in-part
through production of higher excited $c \bar c$ states
and their decay into quarkonia ground state. Since the lifetime of
different sub-threshold quarkonium states is much larger than the
typical life-time of the medium which may be produced in nucleus-nucleus
collisions so their decay occurs almost completely outside the produced
medium. This means that the produced medium can be probed not only by
the ground state quarkonium but also by different excited 
quarkonium states.
Since, different quarkonium states have different sizes (binding energies),
one expects that higher excited states will dissolve at smaller
temperature as compared to the smaller and more tightly bound ground states.
These facts may lead to a sequential suppression pattern in $J/\psi$
yield in nucleus-nucleus collision as the function of the
energy density or number of participants in the collision.
The $J/\psi$ yield could show a
significant suppression even if the energy density of the system is
not enough to melt directly produced $J/\psi$ but
is sufficient to melt the higher resonance states
because they are loosely bound compared to the ground state $J/\psi$.

In nucleus-nucleus collisions, it is known that only about
60\% of the observed $J/\psi$ originate directly in hard collisions while
30\% of them come from the decay of $\chi_c$
and 10\% from the decay of $\psi^\prime$. Hence, the $\pt$-integrated inclusive
survival probability of $J/\psi$ in the QGP becomes~\cite{Sat07,dpal}.
\begin{equation}
\langle S^{{}^{\rm incl}} \rangle = 0.6 {\sdir}_{{}_\psi}
+0.3 {\sdir}_{{}_{\chi_c}}
+0.1 {\sdir}_{{}_{\psi^\prime}}
\end{equation}
The hierarchy of dissociation temperatures in lattice correlator
studies~\cite{Dat04} (Table IV)
thus leads to sequential
suppression pattern with an early suppression of $\psi^\prime$
and $\chi_c$ decay products and much later one
for the direct $J/\psi$ production. However, with our recent results
~\cite{prc-vinet,plb-vinet} (Table II-III) 
employing medium modification to the
full Cornell potential and also results from 
potential model studies~\cite{mocsy_prl} on dissociation temperatures,
all three species will
show essentially almost the same suppression pattern, i.e., the concept
of sequential melting will not have any dramatic effect which
will be seen in the next section. 

\section{Results and discussions}
\begin{table}
\label{table2}
\centering
\caption{Formation time (fm), dissociation temperature
$T_D$ (in units of $T_c$=197 MeV for
a 3-flavor QGP) with the Debye mass in 
the leading-order~\cite{prc-vinet,plb-vinet},
the speed of sound $c_s^2$ and the screening energy density
$\eps$ ($GeV/fm^3$) calculated in SIQGP and ideal EoS
for $J/\psi$, $\psi^\prime$, $\chi_c$ states~\cite{prc-vinet,plb-vinet},
respectively.}
\vspace{3mm}
\begin{tabular}{|l|l|l|l|l|l|l|}
\hline
State &  $\tau_F$  &  $T_D$  &  $c_s^2$(SIQGP)  &  $c_s^2$(Id) & $\epsilon_s$
(SIQGP)
& $\epsilon_s$(Id) \\
\hline\hline
$\jpsi$ &0.89&  1.61 & 0.26 & 1/3 & 17.65 & 21.77 \\
\hline
$\psi'$ & 1.50& 1.16 & 0.24 & 1/3 & 04.51 & 06.53 \\
\hline
$\chi_c$ &2.00& 1.25 & 0.24 & 1/3 &06.15 & 08.47 \\
\hline
\end{tabular}
\end{table}

\begin{table}
\label{table3}
\centering
\caption{Same as Table II but the dissociation temperature
is calculated with the Debye mass in
the lattice parametrized form~\cite{prc-vinet,plb-vinet}.}
\vspace{3mm}
\begin{tabular}{|l|l|l|l|l|l|l|}
\hline
State &  $\tau_F$  &  $T_D$  &  $c_s^2$(SIQGP)  &  $c_s^2$(Id) & $\epsilon_s$
(SIQGP)
& $\epsilon_s$(Id) \\
\hline\hline
$\jpsi$ & 0.89&  1.18 & 0.24 & 1/3 & 4.83 & 7.81 \\
\hline
$\psi'$ & 1.50& 0.85 & 0.18 & 1/3 & 1.21 & 2.89 \\
\hline
$\chi_c$ &2.00& 0.90 & 0.19 & 1/3 & 1.54 & 3.36 \\
\hline
\end{tabular}
\end{table}

\begin{table}
\label{table4}
\centering
\caption{Same as Table II but with different values of
dissociation temperature(s) obtained from lattice studies~\cite{Dat04} 
in units of $T_c$=175 MeV.} 
\vspace{3mm}
\begin{tabular}{|l|l|l|l|l|l|l|}
\hline
State &  $\tau_F$  &  $T_D$  &  $c_s^2$(SIQGP)  &  $c_s^2$(Id) & $\epsilon_s$
(SIQGP) & $\epsilon_s$(Id) \\
\hline\hline
$\jpsi$ &0.89 & 2.10 & 0.27 & 1/3 & 32.85 & 60.83 \\
\hline
$\psi'$ &1.50 & 1.12 & 0.21 & 1/3 & 02.36 & 05.80 \\
\hline
$\chi_c$ &2.00 & 1.16 & 0.22 & 1/3 & 02.74 & 06.53 \\
\hline
\end{tabular}
\end{table}

\begin{table}
\label{table5}
\centering
\caption{Dissociation temperature(s)
$T_D$ (in units of $T_c$=203 MeV for a 2-flavor QGP) 
with the Debye mass in the leading-order~\cite{prc-vinet,plb-vinet} 
of strong coupling.}
\vspace{3mm}
\begin{tabular}{|l|l|l|l|l|l|l|}
\hline
State &  $\tau_F$  &  $T_D$  &  $c_s^2$(SIQGP)  &  $c_s^2$(Id) & $\epsilon_s$
(SIQGP)
& $\epsilon_s$(Id) \\
\hline\hline
$\jpsi$ & 0.89&  1.69 & 0.238 & 1/3 & 17.76 & 23.14 \\
\hline
$\psi'$ & 1.50& 1.21 & 0.172 & 1/3 & 04.07 & 06.70 \\
\hline
$\chi_c$ &2.00& 1.31 & 0.192 & 1/3 & 05.77 & 08.79 \\
\hline
\end{tabular}
\end{table}
There are three time-scales involved  
in the screening scenario of $J/\psi$ suppression in an expanding plasma. 
First one is the screening time, $\tau_s$ as the time available for the
hot and dense system during which $\jsi$'s are suppressed.
Second one is the formation time of $\jpsi$ in the plasma frame
($t_F=\gamma \tau_F$) which depends on the transverse momentum 
by which the $c \bar c$ pairs was produced.
Third one is the cooling rate which depends on the speed of 
sound, $c_s^2$ through the equation of state.
The screening time not only depends upon the 
screening energy density, $\epsilon_s$ but also 
on the speed of sound through EoS. The value of 
$\epsilon_s$ is different for different charmonium 
states and is calculated from the EoS under 
consideration and hence it varies from one EoS to other. 
If $\eps \gtrsim \epsilon_i$, initial energy density, then 
there will be no suppression at all i.e., 
survival probability, $\spt$ is equal to 1.
If the dissociation temperature $T_D$  is higher, 
$\eps$ ($\propto T_D^4$) will also be higher, so system having 
$\epsilon_i > \eps$ will attain $\eps$ quickly. 
Therefore, the system will get less time to kill $\jsi$ in
the deadly region marked by the screening radius, $r_s$ results in
less suppression in the $J/\psi$ yield.  However, for smaller
values of $T_D$, the system will take more time to
reach $\eps$ resulting more suppression. 
\begin{figure*}
\vspace{-1mm}
\includegraphics[scale=.35]{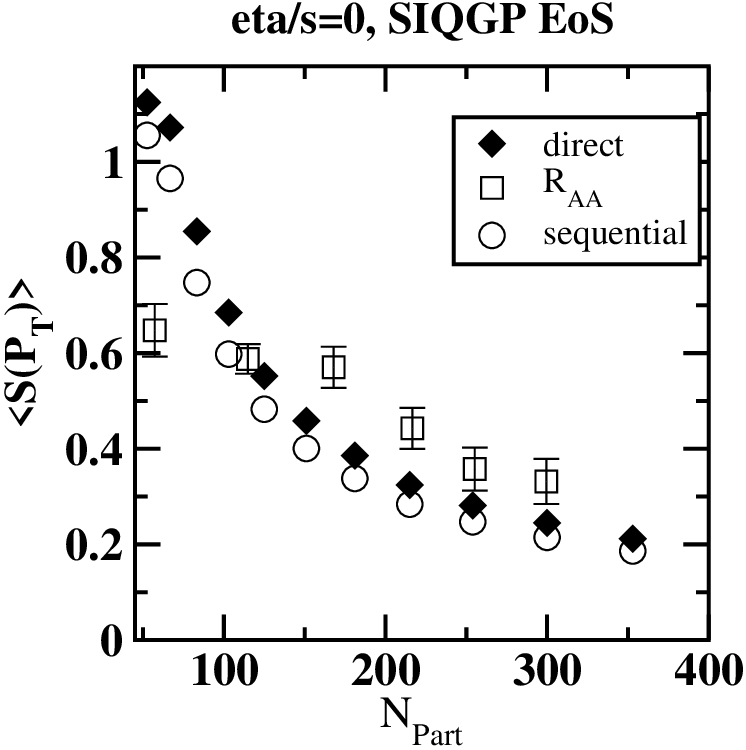} 
\hspace{2mm}
\includegraphics[scale=.35]{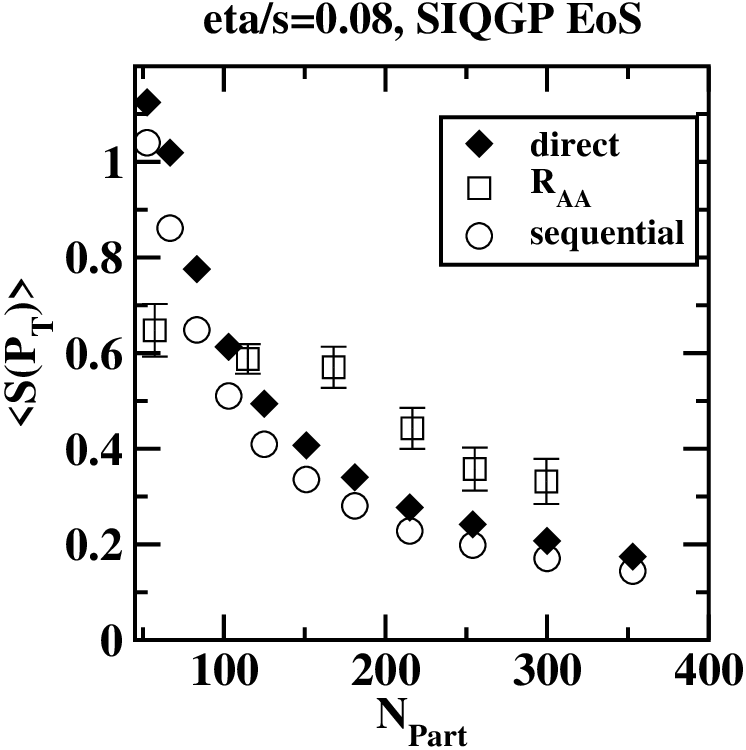}
\hspace{2mm}
\includegraphics[scale=.35]{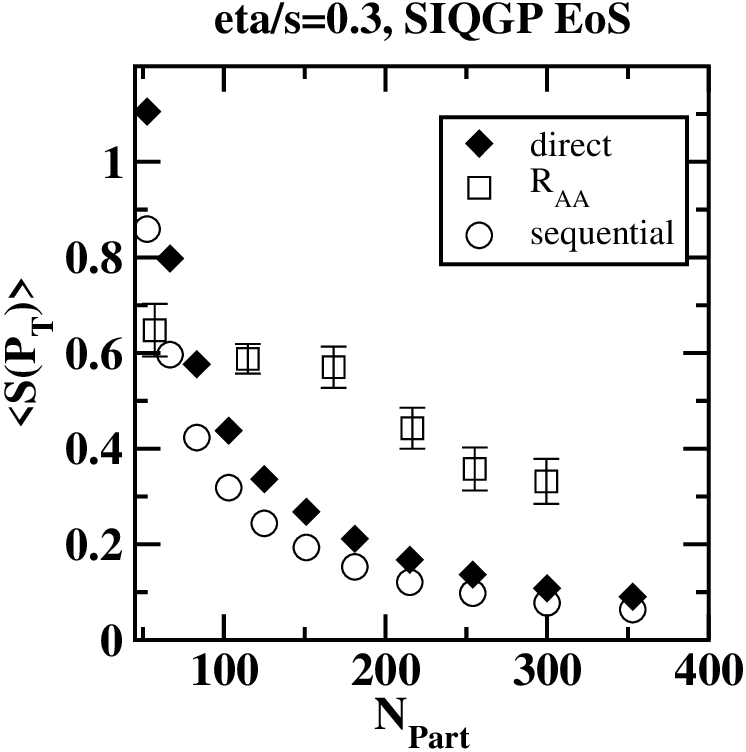}
\vspace{2mm}
\caption{The variation of $\pt$ integrated survival probability
(in the range allowed by invariant $\pt$ spectrum of $J/\psi$ by the
Phenix experiment~\cite{expt}) versus number of participants at mid-rapidity.
The experimental data (the nuclear-modification factor $R_{AA}$)
are shown by the squares with error bars whereas
circles and diamonds represent with ($\sincl$)
without ($\sdir$) sequential melting  using the values of $T_D$'s~
\cite{prc-vinet} and related parameters from Table II using SIQGP equation of
state.}
\vspace{15mm}
\end{figure*}
More precisely, the screening time depends upon (i) the difference 
between the initial energy density $\epsilon_i$ and the
screening energy density $\eps$: the more will be the 
difference the more will be the suppression,
(ii) the speed of sound: the values of $c_s^2$ which are less than
1/3, the rate
of cooling will be slower which, in turn, makes the 
screening time large for a fixed difference in ($\epsilon_i$- $\eps$)
leading to more suppression, and (iii) the $\eta/s$ ratio: if the
ratio is larger then the cooling will be slower, so the system
will take longer to reach $\eps$ resulting the higher value of 
screening time and hence more suppression
compared to $\eta/s=0$. With this physical understanding we analyze 
$\sinpt$ as a function of the number of participants $N_{\mbox{Part}}$ in
an expanding QGP. 

In our analysis, we have employed the
dissociation temperatures for $J/\psi$, $\chi_c$ and $\sip$ recently 
computed by us~\cite{prc-vinet,plb-vinet} in the Table(s) II-III for 3-flavor 
and V for 2 flavor QGP, respectively, where we 
employed an appropriate form of medium modified potential which includes
both non-zero string tension effects beyond $T_c$ and the usual 
screened coulomb term. To compare our predictions with the 
lattice correlator studies, we employ the dissociation 
temperatures (in Table IV) computed from the first principle 
of (lattice) QCD~\cite{Dat04} where a simple screened Coulomb form was 
assumed for the medium-dependent inter-quark potential.
The corresponding values for $\eps$ and $c_s^2$ calculated 
in the strongly-interacting and ideal EoS are also listed in 
the respective Tables. We shall employ these next to study $\sinpt$.

We have shown the variation of $\pt$-integrated survival probability
(in the range allowed by invariant $\pt$ spectrum of $J/\psi$ by the
Phenix experiment~\cite{expt}) with 
$N_{\mbox{Part}}$ at mid-rapidity in Figs.1-8.
The experimental data (the nuclear-modification factor $R_{AA}$)
are shown by the squares with error bars whereas
circles and diamonds represent with ($\sincl$)
without ($\sdir$) sequential melting.
We have employed three values of $\eta/s$ {\em viz.} 0, 0.08, 
and 0.3 in each figure to the see the effects of dissipative terms
in the expansion of the plasma.
In Fig. 1, we find that the survival probability of directly produced 
$\jsi$ is slightly higher than $\sincl$ and is closer to the
the experimental results~\cite{expt}.
For the lower value of $\eta/s$ our predictions are closer to the 
experimental ones.
As the ratio $\eta/s$ is increased, the expansion of the system becomes slower
leading to the 
higher values for the screening time and hence lower values of $\sinpt$
i.e., $J/\psi$'s will be suppressed more.
However, for all the three values of $\eta/s$ ratio, $\sinpt$ 
for both the directly produced and
sequential $J/\psi$ sit below the experimental numbers for 
$N_{\mbox{Part}}\gtrsim 100$. 
\begin{figure*}
\vspace{30mm}
\includegraphics[scale=.35]{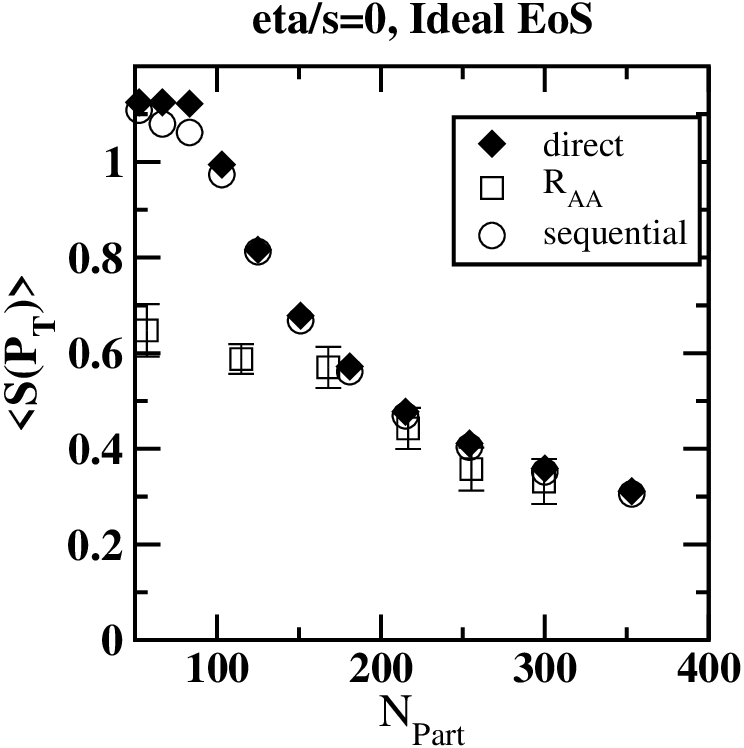} 
\hspace{2mm}
\includegraphics[scale=.35]{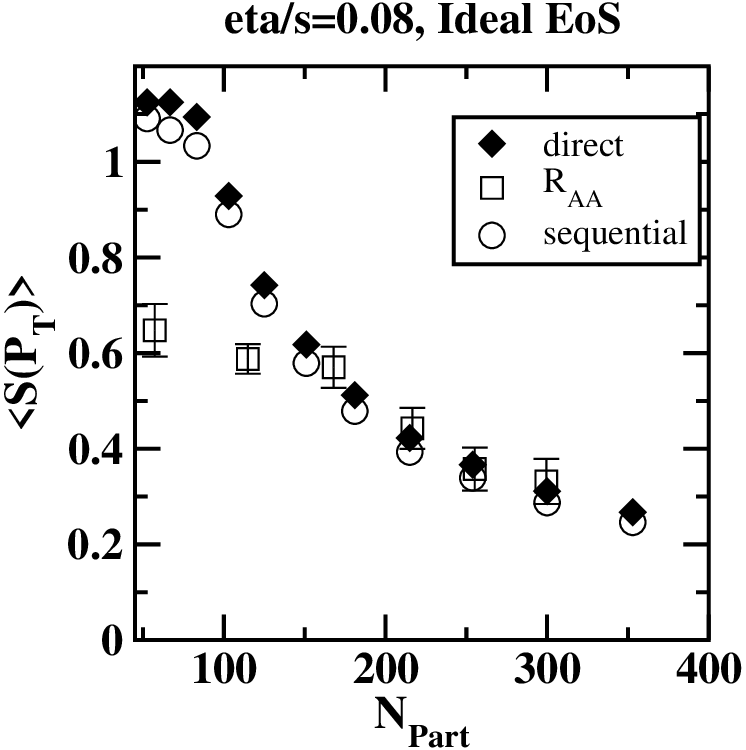}
\hspace{2mm}
\includegraphics[scale=.35]{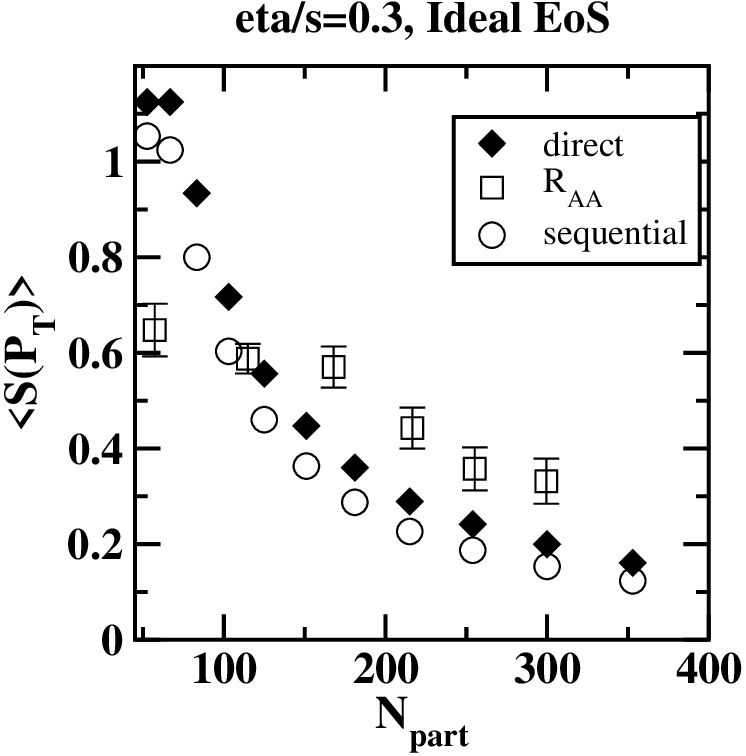}
\vspace{3mm}
\caption{Same as Fig. 1 but the related parameters from the ideal EoS.}
\vspace{35mm}
\end{figure*}

\begin{figure*}
\vspace{5mm}
\includegraphics[scale=.35]{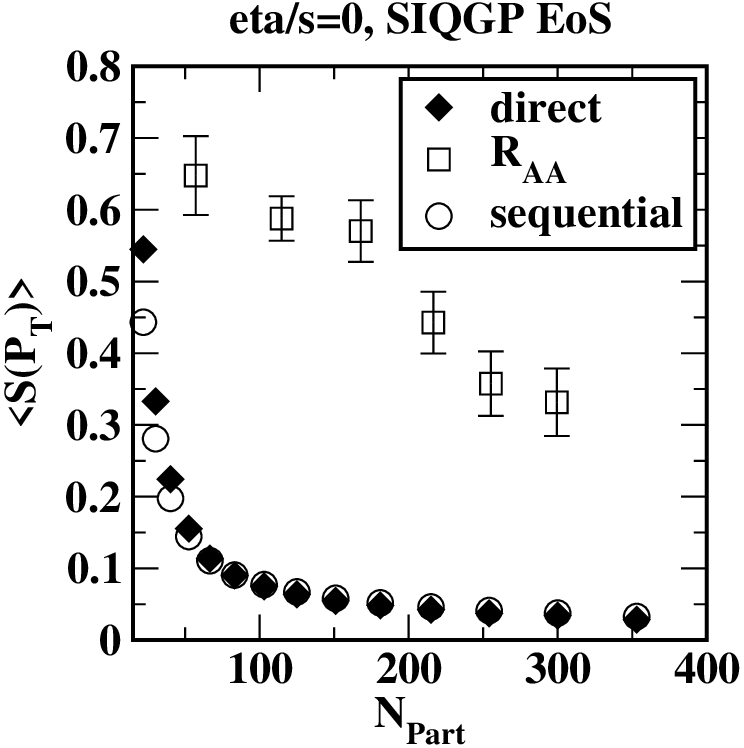}
\hspace{2mm}
\includegraphics[scale=.35]{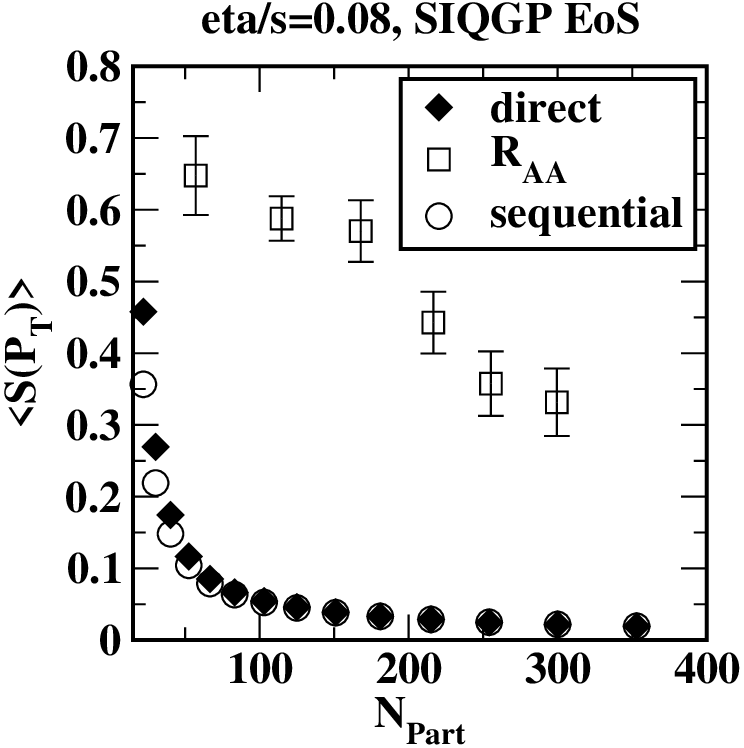}
\hspace{2mm}
\includegraphics[scale=.35]{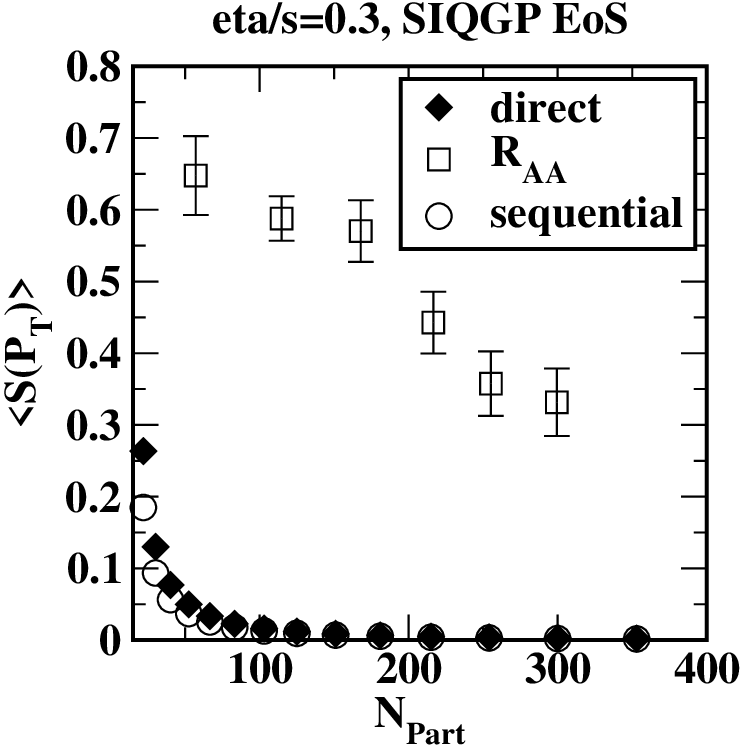}
\vspace{2mm}
\caption{Same as Figure 1 but using 
using the values of $T_D$'s~
\cite{prc-vinet} and related parameters from Table III for the SIQGP EoS.}
\vspace{40mm}
\end{figure*}

\begin{figure*}
\vspace{5mm}
\includegraphics[scale=.35]{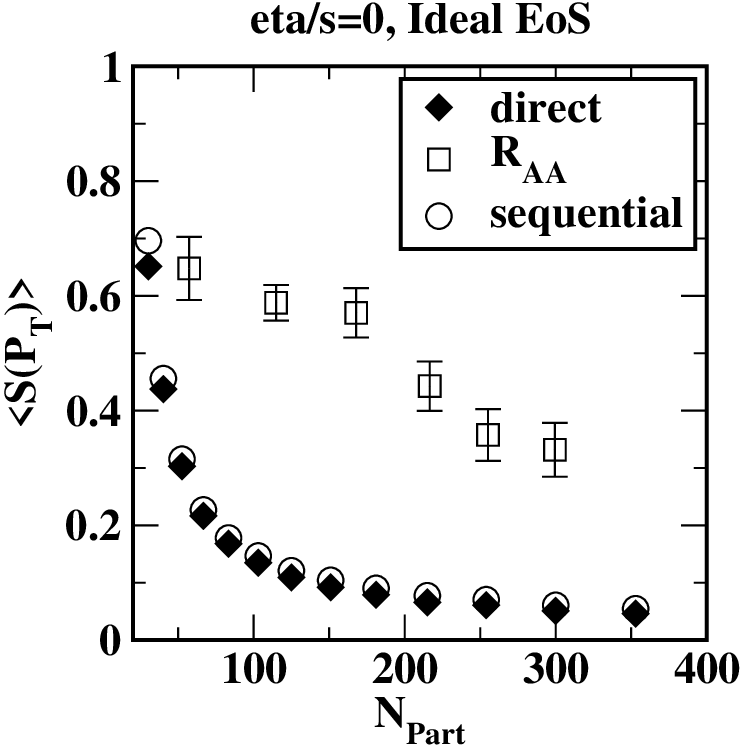}
\hspace{2mm}
\includegraphics[scale=.35]{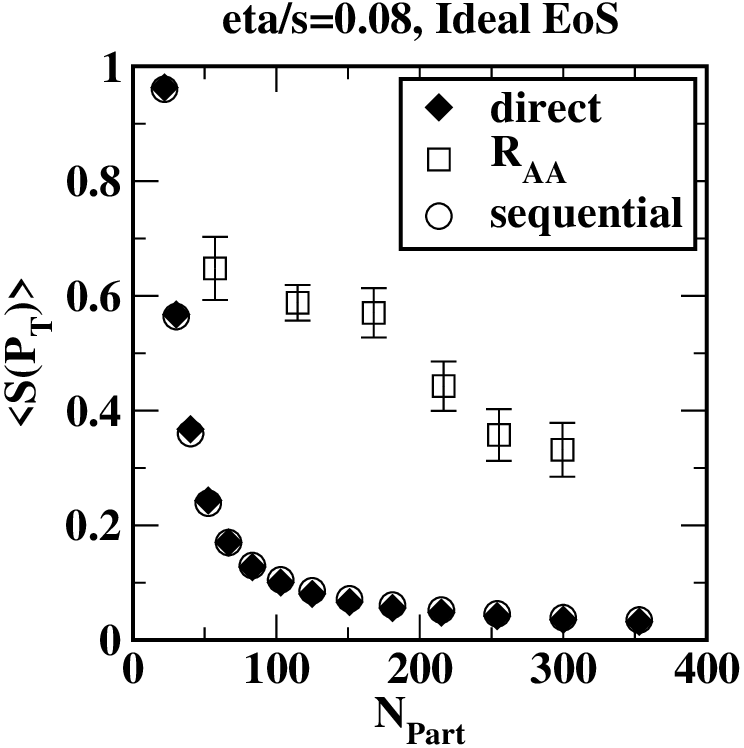}
\hspace{2mm}
\includegraphics[scale=.35]{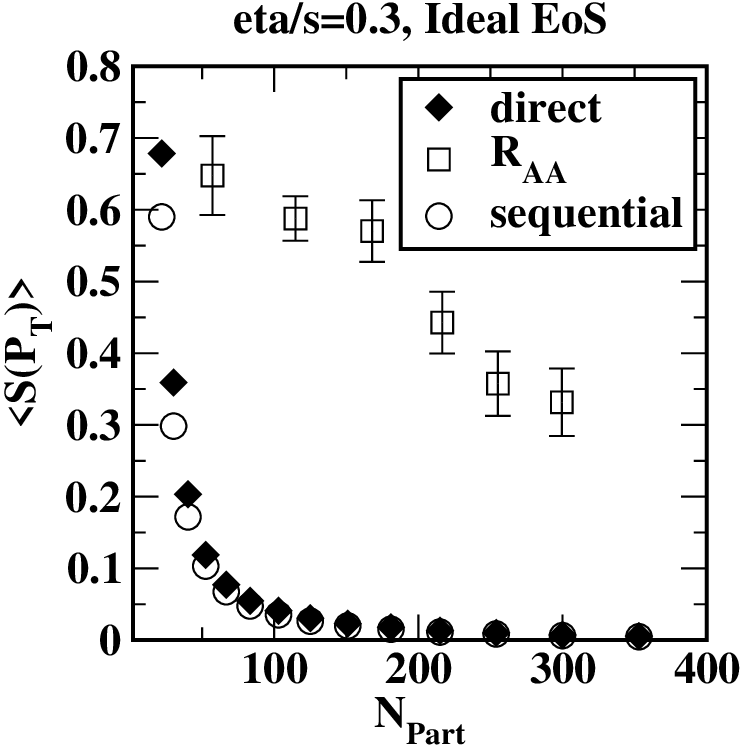}
\vspace{2mm}
\caption{Same as Fig. 3 but employing the ideal EoS}
\vspace{5mm}
\end{figure*}
In Fig.2, we used the same values of dissociation temperatures  as in 
Table II but
the thermodynamic variables {\em viz.} $\eps$, $c_s^2$ etc. have been
calculated in the ideal EoS to see the
sensitivity of the EoS to the plasma dynamics which, in turn, indirectly 
affect suppression in relativistic collision. In this case too, 
the $\eta/s$ dependence of $\sinpt$ pattern remains the same and 
$\sinpt$ is always higher than the corresponding values (with SIQGP) in 
Fig.1. The matching is almost perfect for $\eta/s$=0. 
This can be understood in terms of the sensitivity of $\sinpt$ on
the speed of sound, $c_s^2$  for a fixed difference in 
($\epsilon_i$-$\epsilon_s$)
because cooling of the system with
ideal EoS is much faster compared to the strongly-interacting EoS
so the system will spend less time in the screening region results in
less suppression.

\begin{figure*}
\vspace{-1mm}
\includegraphics[scale=.35]{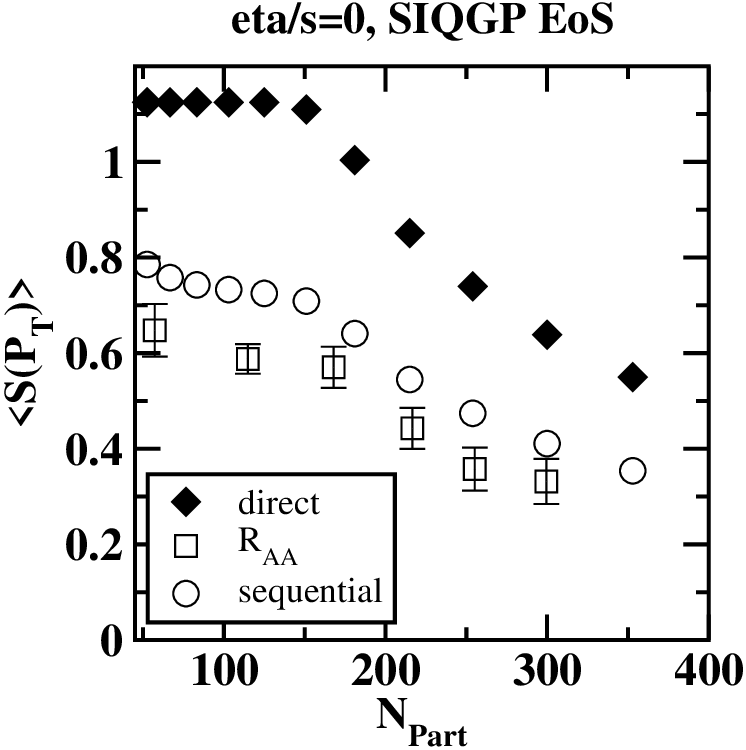} 
\hspace{2mm}
\includegraphics[scale=.35]{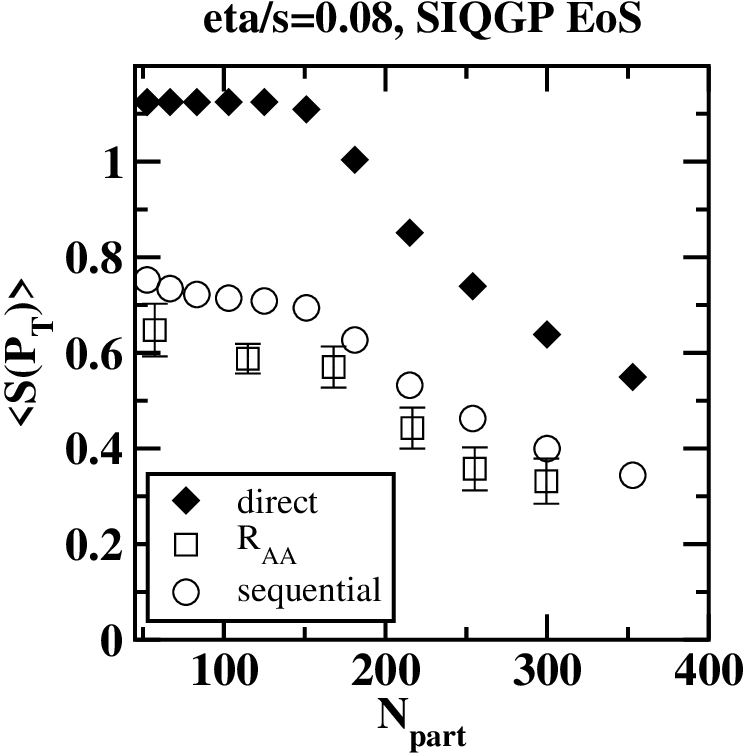}
\hspace{2mm}
\includegraphics[scale=.35]{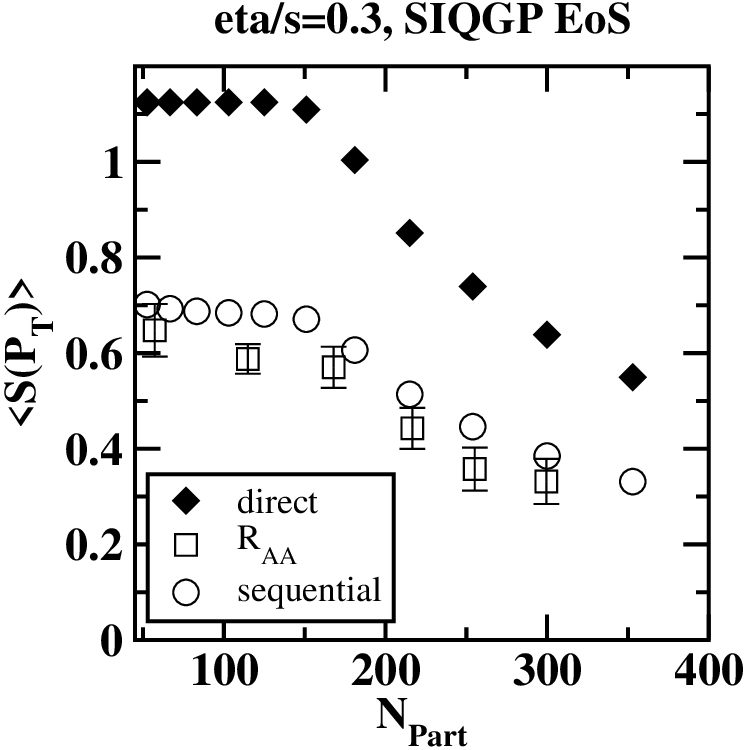}
\vspace{3mm}
\caption{Same as Fig.3 but using the values of $T_D$'s~\cite{Dat04} 
and related parameters from Table IV for the SIQGP EoS.}
\vspace{25mm}
\end{figure*}

Another interesting observation, which is common to both Fig.1 and 2 is 
that as the ratio $\eta/s$ is increased from 0 to 0.3, $\sinpt$ for both
directly and sequential $J/\psi$'s become smaller. To be
more accurate, $\sinpt$ for the sequential decays will be affected 
much causing the suppression stronger and is being separated 
from the $\sinpt$ of the directly produced $\jsi$. 
As the ratio $\eta/s$ is increased from zero, cooling becomes slower
so that $\chi_c$ and $\sip$ show more suppression due to their
smaller value of $\eps$ (smaller $T_D$) compared to $\jsi$ making 
the difference between $\epsilon_i$ and $\eps$ larger. This leads to 
more suppression for $\chi$ and $\sip$'s.

In Figs. 3 and 4, we used the dissociation temperatures 
employing the lattice parametrized form of the Debye mass (in Table III)
which is the same as the values obtained from potential models
but the expansion of the system is taken into 
account through SIQGP and ideal EoS, respectively. 
We again vary $\eta/s$ between $0$ and $0.3$.
Interestingly, in both the cases (Fig.3 and Fig.4) the suppression is
large compared to the previous cases (Fig.1 and Fig.2) and is much
smaller than than the experimental results.
Since the dissociation temperatures are much smaller than
the values in Table II so that the difference ($\epsilon_i-\epsilon_s$) 
becomes large. Therefore, the system will take longer time to 
reach $\epsilon_s$, 
as a result the system have enough time to kill $J/\psi$'s.
The sensitivity to the EoS and the ratio $\eta/s$ is the same 
as in Figs.1 and 2.
As expected, there is more suppression in SIQGP as compared to ideal EoS.

In Figs. 5 and 6, we use the dissociation temperatures from the 
lattice correlator studies~\cite{Dat04} (in Table IV) 
and the expansion of 
the system are taken into account through SIQGP and ideal
EoS, respectively. 
Interestingly, in both Figs. the suppression is much 
smaller compared to the previous cases.
The hierarchy in the dissociation temperatures 
in Table IV led the sequential
suppression to play an important role 
which causes more suppression for the sequentially
produced $J/\psi$'s which is closer to the Phenix results
compared to very less suppression for the directly 
produced $J/\psi$'s.
On the contrary, in the earlier sets on the dissociation 
temperatures the concept
of sequential melting has not any dramatic effect on $\sinpt$.
In this set of dissociation temperatures, the agreement between 
theory and experiment
is better with the SIQGP EoS (Fig.5) compared to ideal EoS (Fig.6)
because cooling is slower in the former EoS and results in more
suppression for the sequential melting. 
In addition to the effect of EoS, the effect of the dissipative
forces in terms of $\eta/s$ becomes prominent because the matching
(with the experiment) is good for higher values of $\eta/s$. This
is due to the fact that the larger value of viscous force makes the expansion
slower which results in more suppression.

To examine the flavor dependence we have plotted the $\pt$-integrated
survival probability for the 2-flavor QGP with the dissociation 
temperatures in Table V in Figs.7, 8. The suppression pattern
remains the same in comparison to Figs.1 and 2. It seems
surprising because of the following reason: As the number of 
flavors in the system decrease, $\epsilon_s$ decrease 
rapidly so that the difference
($\epsilon_i-\epsilon_s$) becomes larger compared to 3-flavor
system. As a result the system will get more time to kill
$J/\psi$'s resulting more suppression.
But it did not happen in the Figs. 7 and 8 because we take into account
the flavor dependence of the dissociation temperature where
the dissociation temperature for the 2-flavor system
is much higher than 3-flavor sysem which compensates 
the decrease in the screening energy density due to the decrease in the 
degrees of freedom. This balance  makes the 
suppression pattern same in Fig.1 and Fig.7
(or Fig.2 and Fig.8). 
\begin{figure*}
\vspace{15mm}
\includegraphics[scale=.35]{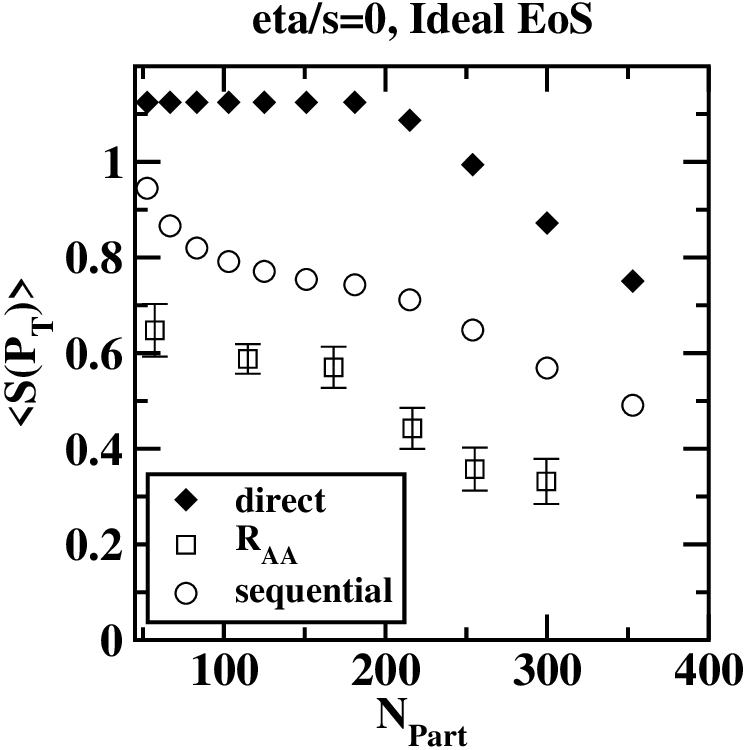} 
\hspace{2mm}
\includegraphics[scale=.35]{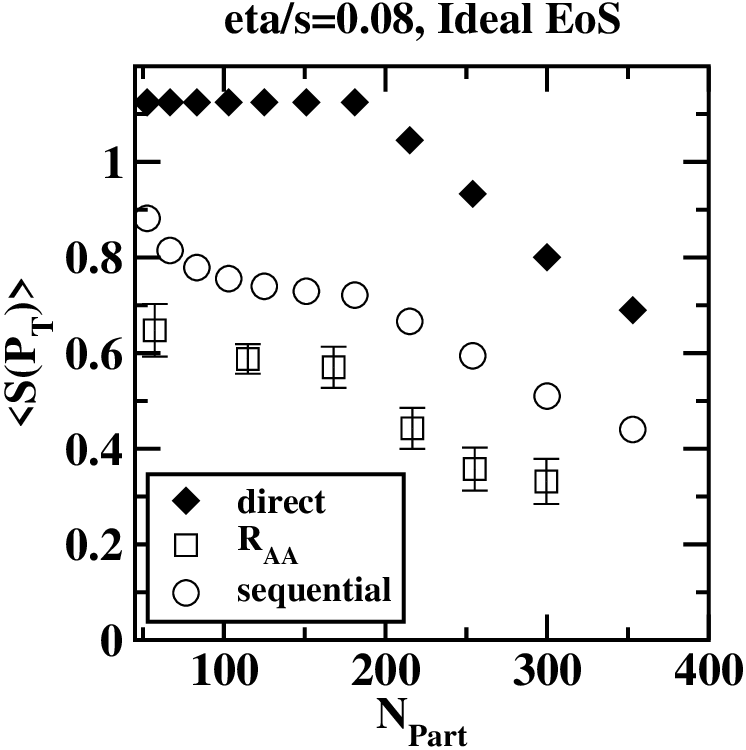}
\hspace{2mm}
\includegraphics[scale=.35]{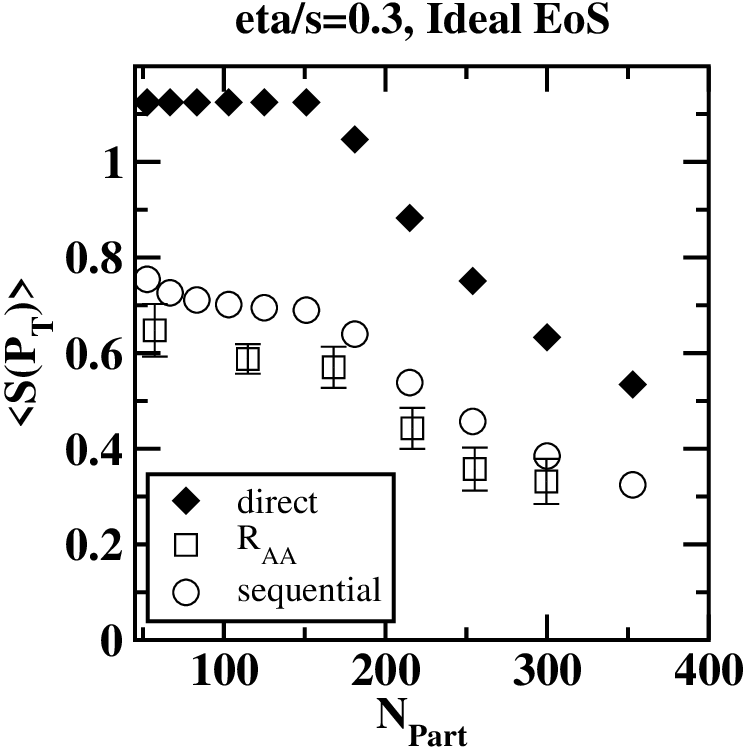}
\vspace{2mm}
\caption{Same as Fig.5 but employing the Ideal EoS.}
\vspace{15mm}
\end{figure*}
\begin{figure*}
\vspace{35mm}
\includegraphics[scale=.35]{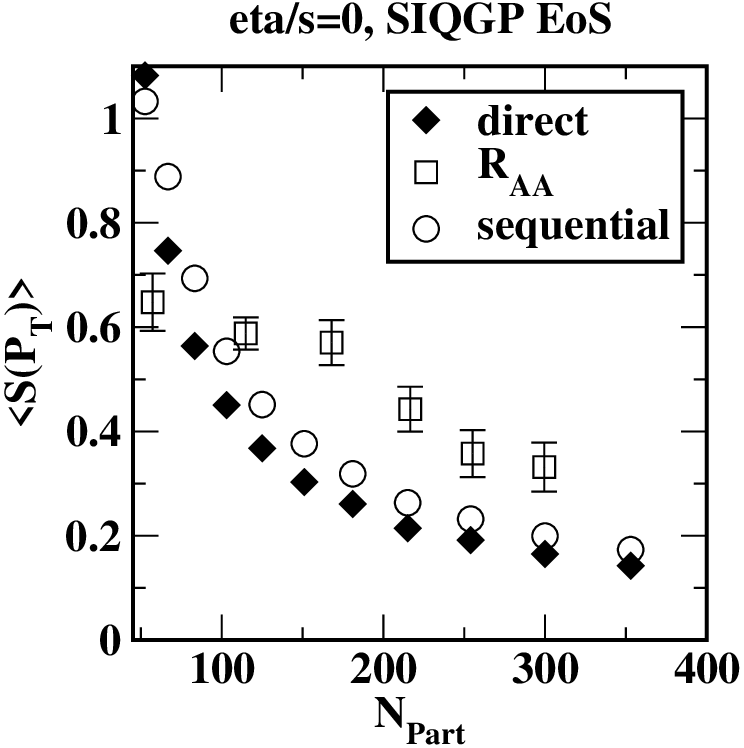}
\hspace{2mm}
\includegraphics[scale=.35]{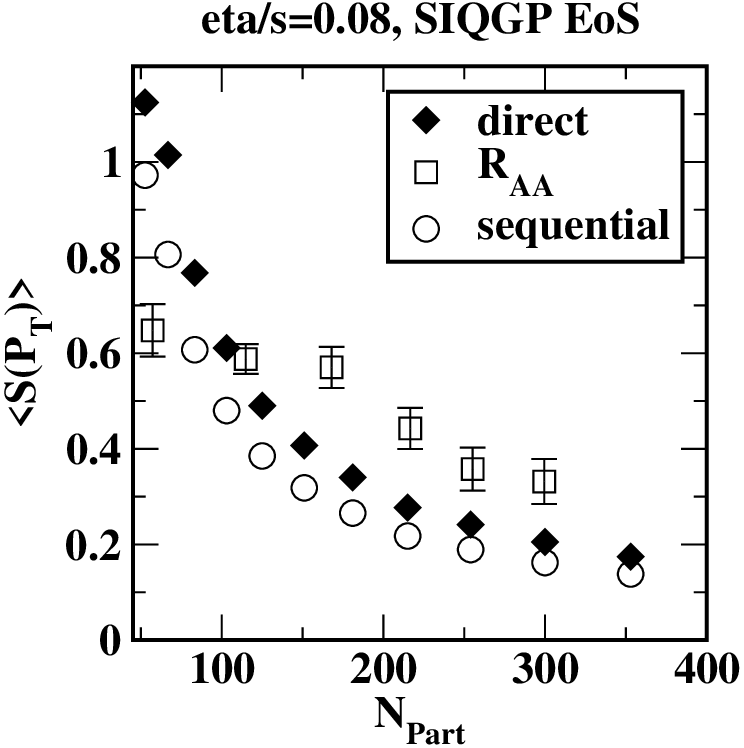}
\hspace{2mm}
\includegraphics[scale=.35]{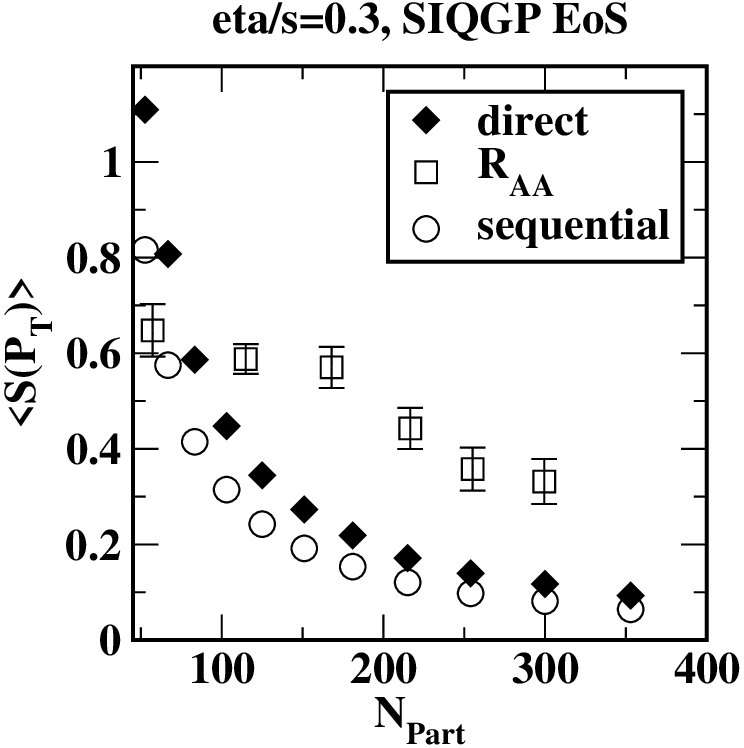}
\vspace{2mm}
\caption{Same as Fig.1 but using the values of $T_D$'s~
\cite{prc-vinet} and related parameters in SIQGP EoS 
from Table V for 2-flavor QGP.}
\vspace{15mm}
\end{figure*}

\begin{figure*}
\vspace{35mm}
\includegraphics[scale=.35]{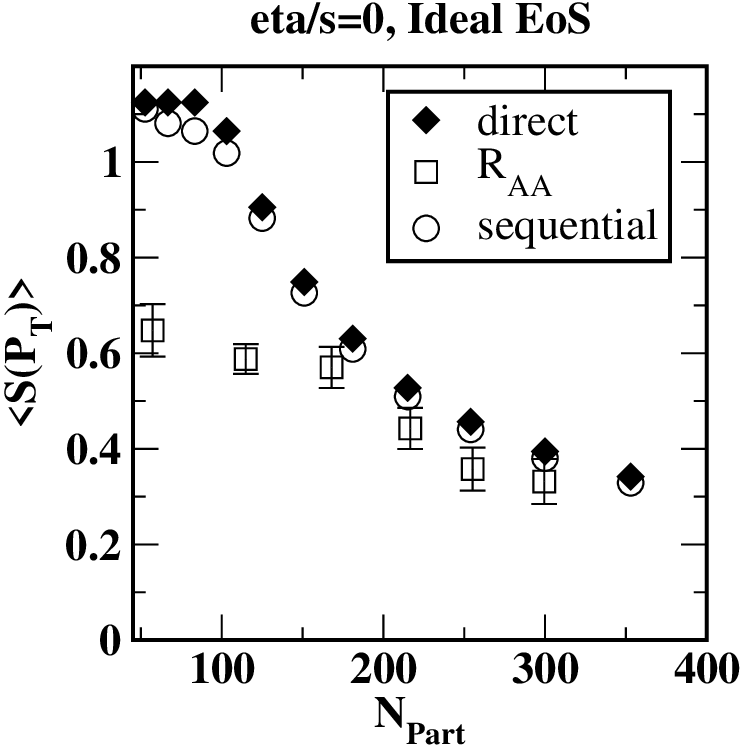}
\hspace{2mm}
\includegraphics[scale=.35]{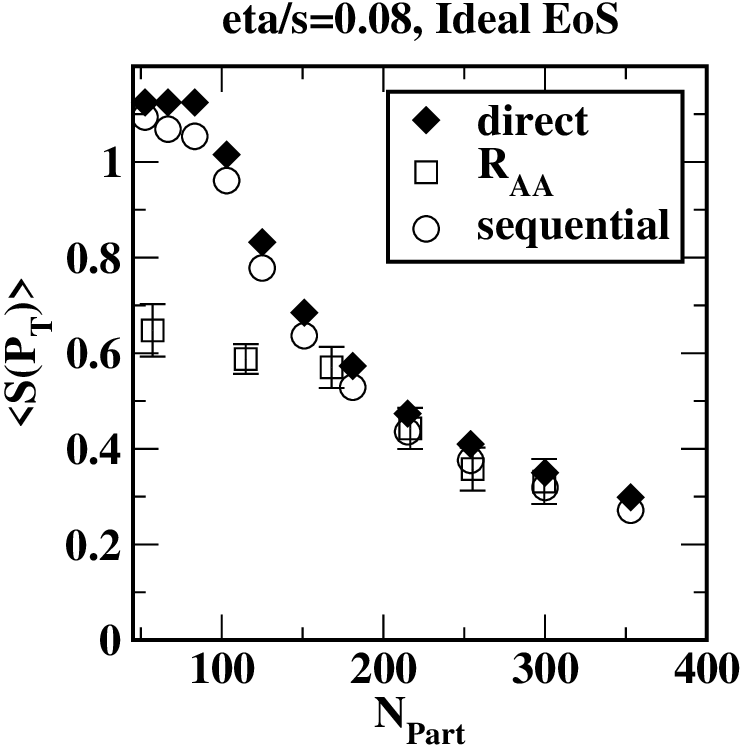}
\hspace{2mm}
\includegraphics[scale=.35]{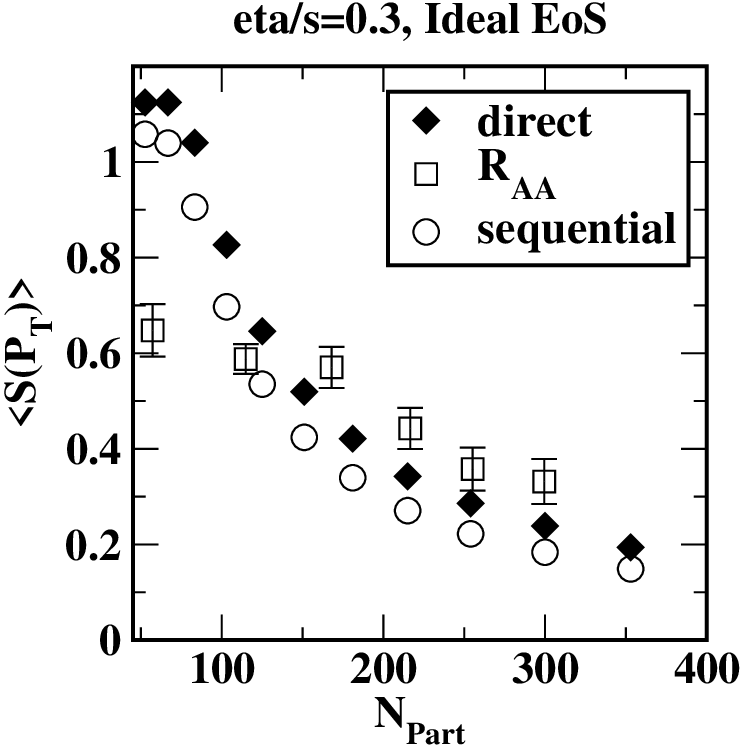}
\vspace{2mm}
\caption{Same as Fig.7 but employing the ideal EoS.}
\vspace{25mm}
\end{figure*}
Let us now compare Fig.1 with Fig.5 and Fig.2 with Fig.6.
It is clear from Figs.5 and 6 that $\sinpt$ for both the directly
and the sequentially produced $J/\psi$ are quite high with 
the higher values of $T_D$'s obtained from the lattice correlator
studies (in Table IV) compared to the values obtained from our model
(in Table II). Since the dissociation temperature for the $J/\psi$
is high so that $\epsilon_s$ is much larger than $\epsilon_i$ up to 
the number of participants 200 resulting no suppression
at all. However, $T_D$'s are not so high for $\chi_c$ and $\psi^\prime$
so that the sequential probability $\langle S^{{}^{\rm incl}} \rangle$
has been suppressed that makes them closer to the experimental 
results for the higher values of $\eta/s$.
However, $\langle S^{{}^{\rm incl}} \rangle$
gets 60\% contribution from the directly 
produced $J/\psi$, it has a dominating contribution on the behavior 
of  sequential $\spt$. This is indeed reflected in Fig.5 and Fig.6. 
To be conclusive, the dissociation temperatures exploited in the present 
work (in Table II) shows better agreement with the experimental 
results from RHIC 
as compared to the spectral function technique calculated either in
the potential model~\cite{mocsy_prl} or in the lattice (temporal) 
correlator~\cite{Dat04}. 

\section{Conclusions and future scope}
In conclusion, we have studied the charmonium suppression in a
longitudinally expanding QGP in the presence of dissipative forces.
We find that presence of dissipative terms in the fluid equation of motion 
slower the expansion rate and eventually lead to the enhanced suppression of 
$J/\psi$. In other words, the presence of viscosity enhances the screening 
time for $J/\psi$ in the SIQGP medium and hence the survival 
probability gets decreased compared to that without the viscous forces.
These conclusions are true for both the directly and 
sequentially produced $J/\psi$. 

In this work, we  have exploited  a recent understanding of 
dissociation of quarkonia in the QGP medium which rely on the 
fact that the transition from the hadronic matter to QGP is a 
crossover not a phase transition in the true sense. We have 
employed the results of~\cite{prc-vinet,plb-vinet} on dissociation
temperatures of various charmonium states. We have employed 
the SIQGP equation of state to estimate the screening 
energy density  and the speed of sound to study the $J/\psi$ yield.
To compare our results with those 
obtained by employing the simple screening picture of quarkonia 
commonly considered in the literature,
we employ dissociation temperatures estimated in lattice 
correlator studies~\cite{Dat04}.
We find that the results on $J/\psi$ survival probability agree
with the Phenix Au-Au data\cite{expt} with the set of dissociation 
temperatures (Table II) obtained with the perturbative result of the
Debye mass. We had shown in our earlier work~\cite{prc-vinet} that
the inclusion of non-perturbative contributions to the Debye mass
lower the dissociation temperatures substantially which looks
unfeasible to compare to the spectral
analysis of lattice temporal correlator of 
mesonic current~\cite{Dat04} which 
finally makes the $J/\psi$ survival probability too small to compare with
the experimental results.
This does not immediately imply that
the non-perturbative effects should be ignored. It is rather interesting 
to investigate the disagreement between the non perturbative result 
obtained with a dimensional-reduction strategy and the Debye mass
arising from the Polyakov-loop correlators.
Only future investigations may throw more light on this
issue.

This leaves an open problem of the agreement between
these two kind of approaches.
This could be partially due to the arbitrariness in the
criteria/definition of the dissociation temperature.
To examine this point we estimated both the upper and lower 
bound on the dissociation temperatures~\cite{prc-vinet,plb-vinet}.
Thus, this study provides us a handle to decipher
the extent up to which non-perturbative effects should
be incorporated into the Debye mass.

Finally, our study does provide a systematic way to analyze the fate of 
charmonium in QGP medium. We have included two important aspects of 
the QGP at RHIC which are based on recent experiment findings:
(i) strongly-interacting picture of QGP and (ii) non-vanishing string 
tension (between $q\bar{q}$) contributions beyond the deconfimenment point.
The first one we have incorporated by 
considering the phenomenological equation of state~\cite{bannur1}
and the speed of sound determined 
from it and the second one is by considering a right criterion for the 
dissociation of quarkonia in QGP. Our attempt is perhaps the first 
one to understand $J/\psi$ suppression systematically in SIQGP. 
It would be of interest to extend the present study by incorporating 
the higher order contributions coming from the viscous forces 
including contributions of the bulk viscosity. These issues will be taken up separately in the near future.

\vspace{5mm}
\noindent{\bf Acknowledgements:} 
VC acknowledges 
Department of Physics, IIT Roorkee for the hospitality under the 
TPSC programme.  VA acknowledges MHRD, New Delhi (India) for 
the financial support.

\end{document}